\documentclass[12pt,fleqn,draft]{article}

\usepackage{a4wide}

\usepackage[intlimits]{amsmath}
\usepackage{amstext,amsfonts,amsbsy,eucal,amssymb}

\usepackage{amsthm}
\usepackage{amscd}

\newcommand{\fR}{\mathcal{R}}
\newcommand{\fT}{\mathcal{T}}
\newcommand{\fP}{\mathcal{P}}
\DeclareMathOperator{\tr}{tr}

\DeclareMathOperator{\fstr}{Str}
\DeclareMathOperator{\ftr}{Tr}
\DeclareMathOperator{\fst}{st}

\title{
Fermionic $R$-Operator and
Algebraic Structure of 1D Hubbard Model:
Its application to quantum transfer matrix
}
\author{Yukiko Umeno  \\
         Department of Physics, Graduate School of Science,\\
         University of Tokyo,\\
         Hongo 7-3-1, Bunkyo-ku, Tokyo 113, Japan}

\date{}
\begin{document}
\maketitle

\vspace{5cm}
\begin{abstract}

The algebraic structure of the 1D Hubbard model is studied by means of 
the fermionic $R$-operator approach. This approach treats the fermion models 
directly in the framework of the quantum inverse scattering method. 
Compared with the graded approach,
this approach has several advantages. 
First, the global properties of the Hamiltonian are naturally reflected 
in the algebraic properties of the fermionic $R$-operator.
We want to note that 
this operator is a local operator acting on fermion Fock spaces. 
In particular, $SO(4)$ symmetry and 
the invariance under the partial particle hole transformation
are discussed.
Second, we can construct a genuinely fermionic quantum 
transfer transfer matrix (QTM) in terms of the fermionic $R$-operator. 
Using the algebraic Bethe Ansatz for the Hubbard model, 
we diagonalize the fermionic QTM and discuss its properties.

\end{abstract} 
\newpage 

\section{Introduction}

Since the discovery of high $T_c$ superconductivity,
strongly correlated electron models
in low dimensionals 
have attracted much more interest.
Among them
the 1D Hubbard model is one of the most interesting solvable models.
The Hamiltonian consists of the hopping term 
and the on-site Coulomb interaction term,
\begin{align}
\mathcal{H} = - \sum_{j=1}^L \sum_{\sigma = \uparrow \downarrow}     
       (c_{j+1 \sigma}^{\dagger} c_{j \sigma} 
                  + c_{j \sigma}^{\dagger} c_{j+1 \sigma})  
       + U \sum_{j=1}^L   
    (n_{j \uparrow}-\frac{1}{2})(n_{j \downarrow}-\frac{1}{2}).
\label{hamiltonian.hubbard}
\end{align}
Lieb and Wu diagonalized the Hamiltonian \cite{lieb}
by using coordinate Bethe ansatz method
under the periodic boundary condition (PBC).
They found that the ground state 
at the finite correlation strength ($U$)
shows property of the insulator in the half-filled case.
and that of conductor in the other case.
Some excited states are studied in \cite{ovchinnikov, coll}.
Takahashi introduced the string hypothesis
to classify the states \cite{takahashi}
and obtained the thermodynamic formulation
The bulk properties, such as the specific heat 
and the magnetic susceptibility, were calculated using this formulation
\cite{kawakami1, usuki}.

In the frame work of the Inverse Scattering method 
\cite{baxter, kulish, sklyanin, korepin, gomez},
the Yang-Baxter equation plays an important role.
Integrability can be discussed in the sense
of an infinite family of conserved currents
which are created from the transfer matrix.
Then the transfer matrix is diagonalized
by using algebraic Bethe ansatz.
At the same time, this procedure can be applied
to diagonalize the quantum transfer matrix
which shows thermodynamic properties.
Shastry
introduced the Jordan-Wigner transformation \cite{jordan}
for the the Hubbard model 
to show the integrability of this model
\cite{shastry1, shastry2, shastry3}.
This corresponding spin model
is called the coupled spin model (the Shastry model).
Mapping to the spin model is a usual method
since the 1D quantum spin model can be easily 
reduced to the 2D classical spin model.
Furthermore the quantum transfer matrix method 
was also applied to the coupled spin model  \cite{kluemper}.
The bulk properties, such as the specific heat 
and the magnetic susceptibility, were successfully calculated. 
They found that the spin-charge separation still remains 
even at the finite temperature.

We have to note, however, that
there exist several differences between the fermion system
and the spin system. 
Because of the non-locality of this transformation,
the PBC for the fermion system
is not equivalent to the PBC for the corresponding spin system.
Even when the system size is infinite,
the correlation function, such as the one particle green function 
$<c^{\dagger}_{j \sigma} c_{k \sigma} > (\sigma=\uparrow, \downarrow)$, 
can not be expressed locally in the spin system
and it is difficult to evaluate the thermodynamic properties
by using quantum transfer matrix method.
Thus another approach is required to investigate local properties.

We introduced recently the fermionic $R$-operator approach, 
where the genuinely fermionic quantum transfer matrix 
can be constructed \cite{sakai}.
The fermionic $R$-operator is constituted from fermion operators.
The integrability is proved for the fermion models
under the PBC \cite{destri, umeno1, umeno2}
and the open boundary condition \cite{umeno3}.
Here the super trace ($\fstr$) and the super transpose ($\fst$) 
for the fermion operators are introduced.
For the spinless fermion model,
the correlation length and the $k_F$ oscillation 
are obtained at a finite temperature \cite{sakai}
through the quantum transfer matrix method. 
This oscillation is a intrinsically fermionic property
and the conformal field theory has already expected it.
The correlation length also agrees with 
the finite temperature correction ($T \rightarrow 0$) \cite{kawakami2}.
For the Hubbard model, we showed the integrability 
using the the fermionic $R$-operator 
in the previous study \cite{umeno2}. 
In this paper, we proceed this study and 
diagonalize the quantum transfer matrix
algebraically to get the thermodynamic properties. 
The algebraic properties are also examined.

We want to mention another attempt to treat fermion models directly,
without mapping to the spin model.
It is based on the graded Yang Baxter equation,
where the quantum space is a fermion fock space
and the auxiliary space is a graded vector space.
The integrability \cite{wadati1, wadati2, wadati3} 
was first proved by Wadati {\it et al.} 
{\it exactly} for the Hubbard model.
The $SO(4)$ symmetry \cite{heilmann, yang1, yang2, pernici, affleck}
of the transfer matrix is shown in \cite{murakami, shiroishi4}.
Ramos and Martins diagonalize the transfer matrix algebraically
\cite{ramos, martins}, which agrees with the coordinate Bethe ansatz method
\cite{lieb}.
Note that several differences exist between this approach and
the fermionic $R$-operator approach.
First, the initial condition of the transfer matrix 
is not a shift operator in the graded approach
and it is difficult to examine the energy momentum.
Second, the fermionic $R$-operator approach is more applicable
when we apply the quantum transfer matrix method
and diagonalize it algebraically.
In this approach, both the quantum space and the auxiliary space
are the fermion fock spaces.
On the other hand, 
the quantum transfer matrix acts on the auxiliary space
while the transfer matrix acts on the quantum space.
Then 
both transfer matrices have a similar algebraic property
and can be diagonalized almost in the same way.

This paper is constituted as follows.
In the next section, we review the fermionic $R$-operator
and the integrability of this model.
In \S 3, the properties of the fermionic $R$-operator
and the relation with global symmetry is clarified.
In \S 4, we diagonalize the transfer matrix 
by means of algebraic BA approach.
Then we generalize the eigenvalue of the vacuum state
which is useful in diagonalizing the quantum transfer matrix
in \S 5.
The twisted boundary condition is also considered.
In \S 5, we show that 
the monodromy matrix for the quantum transfer matrix
is intertwined by the fermionic $R$-operator
just in the same as that for the transfer matrix.
Then we diagonalize it algebraically
and compare with \cite{kluemper}.
The last section is devoted to concluding remarks.

\section{Fermionic $R$-operator and Yang Baxter equation}
\setcounter{equation}{0}
\renewcommand{\theequation}{2.\arabic{equation}}
In this section we review the fermionic $R$-operator approach
for the 1D Hubbard model (\ref{hamiltonian.hubbard}).
The integrability can be discussed by using the Yang-Baxter equation
\cite{umeno2},
\begin{align}
\fR_{12}(u_1,u_2) \fR_{13}(u_1,u_3) \fR_{23}(u_2,u_3)
= \fR_{23}(u_2,u_3) \fR_{13}(u_1,u_3) \fR_{12}(u_1,u_2).
\label{ybe}
\end{align}
The monodromy operator $\fT_a (u)$
can be intertwined by the fermionic $R$-operator,
\begin{align}
&\fR_{ab}(u_a,u_b) \fT_{a}(u_a) \fT_{b}(u_b)
= \fT_{b}(u_b) \fR_{a}(u_a) \fR_{ab}(u_a,u_b),
\label{glybrel}\\
&\fT_{a}(u)=\fR_{aL}(u,0) \cdots \fR_{a1}(u,0)
\label{monodromy2}
\end{align}
which leads to the commutativity of the transfer matrix $\tau(u)$,
\begin{align}
&[\tau(u),\tau(v)]=0, \ \
\tau(u) = \fstr_a \fT_a(u).
\label{transfermatrix}
\end{align}
Here $\fstr$ is defined as (\ref{fstr})(\ref{tm}).
And the commuting family
can be produced from the transfer matrix,  
\begin{align}
[I^{(m)}, I^{(n)}] = 0,  \ \ 
I^{(n)} = \dfrac{\rm{d}^n}{\rm{d}u^n}
 \{ \tau(0)^{(-1)} \tau(u) \} |_{u=0},
\label{conservedcurrent}
\end{align}
where $I^{(1)}$ is the Hamiltonian.
The fermionic $R$-operator which satisfies (\ref{ybe}) 
\cite{shiroishi1, shiroishi2, shiroishi3, umeno2} 
is given by
\begin{align}
\fR^{\rm h}_{12}(u,v)  &=
\fR^{(\uparrow)}_{12}(u-v) 
\fR^{(\downarrow)}_{12}(u-v)
+  \dfrac{\cos (u-v)}{\cos (u+v)} \tanh (h(u) - h(v)) \nonumber \\
& \times \fR^{(\uparrow)}_{12}(u+v) 
\fR^{(\downarrow)}_{12}(u+v)
(2n_{1 \uparrow}-1)(2n_{1 \downarrow}-1),
\label{fermionicRhubbard}\\
\fR^{(\sigma)}_{12}(u)=&
a(u)(-n_{1 \sigma}n_{2 \sigma}+(1-n_{1 \sigma})(1-n_{2 \sigma}))
-b(u)(n_{1 \sigma}(1-n_{2 \sigma})+(1-n_{1 \sigma})n_{2 \sigma})
\nonumber\\
&+c(u)(c^{\dagger}_{1 \sigma}c_{2 \sigma}+c^{\dagger}_{2 \sigma}c_{1 \sigma}
), \ \ \ \ (\sigma=\uparrow, \downarrow)\\
\rm{where} \ \
&a(u)=\cos u, \ \ b(u)=\sin u, \ \  c(u)=1, \ \ 
\frac{\sinh 2h(u)}{\sin 2u}=\frac{U}{4},
\end{align}
which is slightly different 
from the one in \cite{umeno2} because of the different JW trans.
Note that the initial condition for this $R$-operator (u=v=0)
gives the permutation operator ($\fP$),
which means that the initial condition for 
the transfer matrix $\tau(0)$ is a left shift operator,
\begin{align}
&\fP_{jk}c_{j \sigma} = c_{k \sigma} \fP_{jk}, \ \ 
\fP_{jk} c_{k \sigma} = c_{j \sigma} \fP_{jk} \ \ \rm{and} \ \
\fP_{jk}^2=1,
\label{permutation}\\
&\tau(0) c_{j \sigma} = c_{j+1 \sigma} \tau(0), \ \ 
\rm{with} \ \ \rm{PBC} \ \ ( c_{L+j}=c_j),
\label{shift}
\end{align}
The Hamiltonian (\ref{hamiltonian.hubbard}) is expressed as
\begin{align}
&H =\tau(0)^{-1} \frac{d}{du} \tau(u)|_{u=0} 
=\sum_{j=1}^L \fP_{j,j+1} \frac{d}{du} \fR_{j,j+1}(u,0)|_{u=0}.
\end{align}
This is the fermionic $R$-operator approach for the integrability.
We see that (\ref{glybrel})(\ref{monodromy2}) can be generalized \cite{umeno2}
to $\fT_{a}(u)=\fR_{aL}(u,u_0) \cdots \fR_{a1}(u,u_0)$
where $\fR_{jk}(u_0,u_0)=\fP_{jk}$. 
Then the generalized Hamiltonian is obtained.
It was generally believed that
we demand the deference property $\fR(u,v)=\fR(u-v)$
to get the boost operator. The Hubbard model does not have this property. 
Recently J. Links {\it et al.} found the boost operator
by using this generalized formula \cite{links}.

\section{${SO(4)}$ Symmetry and Partial Particle-Hole Transformation}
\setcounter{equation}{0}
\renewcommand{\theequation}{3.\arabic{equation}}
In this section we discuss the two important symmetries 
of the Hubbard model by means of the fermionic ${R}$-operator.
The first one is the ${SO(4)}$ symmetry which consists of 
the spin-${SU(2)}$ and the charge-${SU(2)}$ (${\eta}$-pairing ${SU(2)}$). 
The generators of the spin-${SU(2)}$ and the charge-${SU(2)}$ 
are given by
\begin{align} 
&S^{x} = \frac{1}{2} \sum_{j=1}^{L}  
(c_{j\uparrow}^{\dagger} c_{j\downarrow} 
+ c_{j\downarrow}^{\dagger} c_{j\uparrow}), \ \ 
S^{y} = \frac{1}{2 {\rm i}} \sum_{j=1}^{L}  
(c_{j\uparrow}^{\dagger} c_{j\downarrow} 
- c_{j\downarrow}^{\dagger} c_{j\uparrow}), \ \ 
S^{z} = \frac{1}{2} \sum_{j=1}^{L}  
(n_{j\uparrow} - n_{j\downarrow}), 
\label{eq.spin-su(2)}\\
& \eta^{x} = \frac{1}{2} \sum_{j=1}^{L} 
(-1)^{j} ( c_{j\uparrow}^{\dagger} c_{j\downarrow}^{\dagger} 
+ c_{j\downarrow} c_{j\uparrow} ), \ \
\eta^{y} = \frac{1}{2 {\rm i}} \sum_{j=1}^{L} 
(-1)^{j} ( c_{j\uparrow}^{\dagger} c_{j\downarrow}^{\dagger} 
- c_{j\downarrow} c_{j\uparrow} ), \nonumber \\ 
& \eta^{z} = \frac{1}{2} \sum_{j=1}^{L}  
(n_{j\uparrow}+n_{j\downarrow}-1). 
\label{eq.charge-su(2)}
\end{align} 
These generators constitute the Lie algebra of ${SO(4)}$, 
i.e., ${so(4) \equiv su(2) \oplus su(2)}$. 
It is well known that the six generators 
(\ref{eq.spin-su(2)}) and (\ref{eq.charge-su(2)}) 
commute with the Hubbard Hamiltonian (\ref{hamiltonian.hubbard}),
\begin{equation}
\left[ H, S^{\alpha} \right] =  \left[ H, \eta^{\alpha} \right] = 0, 
\ \ \ \  (\alpha = x, y, z)
\end{equation}
if we assume the even number of sites and the PBC. 

The second one is the so called partial particle hole transformation,
\begin{equation}
  c_{j \uparrow} \rightarrow c_{j \uparrow}, \ \ \ \ 
  c_{j \downarrow} \rightarrow (-1)^{j} c_{j \downarrow}^{\dagger},
  \ \ \ \ U \rightarrow -U.
\label{eq.pphtransf}
\end{equation} 
It is also well known that the Hubbard Hamiltonian is invariant 
under the partial particle-hole transformation, 
while the generators ${S^{\alpha}}$ and ${\eta^{\alpha}}$ 
are exchanged to each other. 

We discuss the above two symmetries 
in terms of the fermionic ${R}$-operator. 
Here we omit some calculations,
which we will show in Appendix A.
The Lie algebra ${so(4)}$ symmetry of the 
fermionic ${R}$-operator was already discussed in \cite{umeno2} , 
which is represented by the relations, 
\begin{align}
& \left[ \fR_{jk}(u,v), 
S_{j}^{\alpha} + S_{k}^{\alpha} \right]=0, \ \ \ \   
(\alpha = x,y,z) \label{eq.comm1} \\
& \left[ \fR_{jk}(u,v), 
\eta_{j}^{z} + \eta_{k}^{z} \right]=0,\label{eq.comm2} \\
& \left\{ \fR_{jk}(u,v), 
\eta_{j}^{\alpha} - \eta_{k}^{\alpha} \right\}=0. \ \ \ \ 
(\alpha = x,y) \label{eq.comm3} 
\end{align}
Here we have introduced the local generators  as
\begin{align}  
& S_{j}^{x} = \frac{1}{2} (c_{j\uparrow}^{\dagger} c_{j\downarrow} 
+ c_{j\downarrow}^{\dagger} c_{j\uparrow}), \ \  
S_{j}^{y} = \frac{1}{2 {\rm i}} (c_{j\uparrow}^{\dagger} c_{j\downarrow} 
- c_{j\downarrow}^{\dagger} c_{j\uparrow}), \ \ 
S_{j}^{3} = \frac{1}{2} (n_{j\uparrow} - n_{j\downarrow}), \nonumber \\ 
& \eta_j^{x} = \frac{1}{2} 
( c_{j\uparrow}^{\dagger} c_{j\downarrow}^{\dagger} 
+ c_{j\downarrow} c_{j\uparrow} ), \ \
 \eta_j^{y} = \frac{1}{2 {\rm i}} 
( c_{j\uparrow}^{\dagger} c_{j\downarrow}^{\dagger} 
+ c_{j\downarrow} c_{j\uparrow} ), \ \ 
\eta_j^{z} = \frac{1}{2} (n_{j\uparrow}+n_{j\downarrow}-1).   
\nonumber
\end{align}
The relations (\ref{eq.comm1})--(\ref{eq.comm3}) can be 
extended to the those for the transfer matrix, 
from which it is possible to prove the ${so(4)}$ symmetry of 
the local conserved currents. 

We shall generalize the symmetry relations 
(\ref{eq.comm1})--(\ref{eq.comm3}) into exponentiated forms, 
which corresponds to the generalization of the Lie algebra ${so(4)}$ symmetry 
into Lie group ${SO(4)}$ symmetry.  By direct calculation, 
we have found the following identities,
\begin{align}
& \left[ \fR_{jk}(u,v),  \exp ({\rm i} \theta S_j^{\alpha}) 
\exp ({\rm i} \theta S_k^{\alpha}) \right] = 0 , \ \ \ \ (\alpha = x,y,z) 
\label{eq.exp1} \\
& \left[ \fR_{jk}(u,v),  \exp ({\rm i} \theta \eta_j^{z}) 
\exp ({\rm i} \theta \eta_k^{z}) \right] = 0 , \label{eq.exp2} \\
&  \fR_{jk}(u,v) \exp({\rm i} \theta \eta_j^{\alpha}) 
\exp(-{\rm i} \theta \eta_j^{\alpha}) = \exp(- {\rm i} \theta \eta_j^{\alpha})
\exp({\rm i} \theta \eta_k^{\alpha}) \fR_{jk}(u,v), \ \ \ \ (\alpha = x,y) 
\nonumber \\ \label{eq.exp3} 
\end{align}
Here ${\theta}$ is an arbitrary (real) parameter. 
From the symmetry relations (\ref{eq.exp1})--(\ref{eq.exp3})
for the fermionic ${R}$-operator, 
we can derive the following identity for the transfer matrix, (see Appendix A) 
\begin{align}
&\exp(- {\rm i} \theta S^{\alpha}) \tau(u,u_0) 
\exp( {\rm i} \theta S^{\alpha}) 
=  \tau(u,u_0), \ \ \ \ (\alpha = x,y,z) 
\label{eq.trexp1} \\
&\exp(- {\rm i} \theta \eta^{z}) \tau(u,u_0) \exp( {\rm i} \theta \eta^{z}) 
= \tau(u.u_0), 
\label{eq.trexp2}\\
&  \exp( {\rm i} \theta \eta^{\alpha}) \tau(u,u_0) 
\exp( {\rm i} \theta \eta^{\alpha}) 
= \tau(u,u_0). \ \ \ \ (\alpha = x,y) 
\label{eq.trexp3}
\end{align}
When we think about the local conserved currents ($I^{(m)}$)
as was discussed in (\ref{conservedcurrent}), 
we consider 
the product ${\tau^{-1}(u_0,u_0) \tau(u,u_0)}$
and it satisfies the relation
\begin{equation}
{\rm e}^{- {\rm i} \theta X} \tau^{-1}(u_0,u_0) \tau(u,u_0) {\rm e}^{ {\rm i} \theta X}
 = \tau^{-1}(u_0,u_0) \tau(u,u_0),
\label{so4sym}
\end{equation}
where ${X}$ is any generator of ${SO(4)}$, i.e., 
${S^{\alpha}}$ or ${\eta^{\alpha}}$ $(\alpha=x, y, z)$. 
Namely the local conserved currents which contain the Hamiltonian
are therefore invariant under the ${SO(4)}$ rotation.

Next we consider the partial particle-hole transformation 
(\ref{eq.pphtransf}). 
The relevant relations for the fermionic ${R}$-operator are found to be
\begin{align}
& \fR_{jk}(u,v;U) V_j \bar{V}_k= V_{k} \bar{V}_j
\fR_{jk}(u,v;-U), \label{eq.ph1}\\
& \fR_{jk}(u,v;U) \bar{V}_j V_k= \bar{V}_{k} V_j
\fR_{jk}(u,v;-U), 
\label{eq.ph2}\\
& \rm{where} \ \
 V_{j} = ( 1 - 2 n_{j\uparrow} ) ( c_{j\downarrow}^{\dagger} 
+ c_{j\downarrow} ), \ \ 
 \bar{V}_{j} = {\rm i} ( 1 - 2 n_{j\uparrow} ) 
(c_{j\downarrow}^{\dagger} - c_{j\downarrow} ) 
= {\rm i} (2 n_{j\downarrow} -1) V_{j}.
\end{align}
Here we write the ${U}$-dependence of the ${R}$-operator explicitly. 
Note that ${V_{j}}$ and ${\bar{V}_{j}}$ are Grassmann odd operators 
and ${ V_j^2 = \bar{V}_j^2 = 1}$.  The operators ${V_j}$ and ${\bar{V}_j}$ 
induce the following transformation for the fermion operators, 
\begin{align}
& V_j c_{j \uparrow} V_j = c_{j \uparrow}, \ \ 
V_j c_{j \uparrow}^{\dagger} V_j = c_{j \uparrow}^{\dagger}, \ \ 
V_j c_{j \downarrow} V_j = c_{j \downarrow}^{\dagger}, \ \ 
V_j c_{j \downarrow}^{\dagger} V_j = c_{j \downarrow}, \nonumber \\
&  \bar{V}_j c_{j \uparrow} \bar{V}_j = c_{j \uparrow}, \ \ 
\bar{V}_j c_{j \uparrow}^{\dagger} \bar{V}_j = c_{j \uparrow}^{\dagger}, \ \
\bar{V}_j c_{j \downarrow} \bar{V}_j = - c_{j \downarrow}^{\dagger}, \ \ 
\bar{V}_j c_{j \downarrow}^{\dagger} \bar{V}_j = - c_{j \downarrow}. 
\label{eq.Vtransf}
\end{align}
Now let us define the operator ${V}$ by
\begin{align}
  V  = \prod_{k=1}^{L/2}
V_{2 k} \bar{V}_{2k -1}
   = V_{L} \bar{V}_{L-1} \cdots V_{2} \bar{V}_1, 
\end{align}
From the property (\ref{eq.Vtransf}), one can see that 
the transformation for 
an arbitrary operator ${X(\{ c_{j \sigma} \}, U)}$ 
\begin{equation}
X(\{ c_{j \sigma} \}, U) \longrightarrow V^{-1}  
X(\{ c_{j \sigma} \}, -U) V
\end{equation}
is nothing but the partial particle-hole transformation 
(\ref{eq.pphtransf}).
From the properties of the fermionic $R$-operator
(\ref{eq.ph1})(\ref{eq.ph2})
the product ${\tau^{-1}(u_0,u_0) \tau(u,u_0)}$  
is invariant under the partial particle-hole transformation,
(see Appendix A)
\begin{equation}
V^{-1} \tau^{-1}(u_0,u_0) \tau(u,u_0;-U) V =  \tau^{-1}(u_0,u_0) \tau(u,u_0;U).
\label{phinv}
\end{equation}
Then we can conclude that all the local conserved currents are 
invariant under the partial particle-hole transformation (\ref{eq.pphtransf}).

\section{Algebraic Bethe Ansatz}
\setcounter{equation}{0}
\renewcommand{\theequation}{4.\arabic{equation}}
In this section, we propose the brief note of the algebraic bethe ansatz.
As Ramos and Martins did in the graded case \cite{ramos},
we construct the eigenstates algebraically
although a little difference exists.
The eigenvalue of the transfer matrix and 
the nested bethe ansatz equation are obtained.
Then we apply
the generalized eigenvalue of the vacuum state,
which we will use in \S 5.
Twisted boundary condition is also discussed.

We will diagonalize the transfer matrix (\ref{fstr})(\ref{tm}), 
\begin{align}
\tau(u) = \fstr_a \fT_a(u) =D_{11}(u)+D_{22}(u)-A_{11}(u)-A_{22}(u).
\end{align} 
The eigenstate is constructed algebraically.
We have five creation operators 
($C_{21}$, $C_{22}$, $D_{21}$, $B_{11}$ and $B_{21}$) 
defined in (\ref{monodromy}). 
By using three creation operators ($C_{21}$, $C_{22}$, $D_{21}$) 
and taking liner combination of these products,
we define the n-particle eigenstate by
\begin{align}
|\Phi_n & (\lambda_1, \cdots ,\lambda_n) \rangle 
=\vec{\Phi}_n(\lambda_1, \cdots ,\lambda_n) 
.\vec{\mathcal{F}} | 0 \rangle,
\label{eigenvecter}
\\
\vec{\Phi}_n & (\lambda_1, \cdots ,\lambda_n)
= \vec{C}(\lambda_1) \otimes \vec{\Phi}_{n-1}(\lambda_2, \cdots , \lambda_n)
\nonumber\\
&+ \Sigma_{j=2}^n
[ \vec{\xi} \otimes D_{21}(\lambda_1) 
\vec{\Phi}_{n-2}(\lambda_2, \cdots ,\check{\lambda}_j, \cdots ,\lambda_n)
D_{22}(\lambda_j) ]
.\hat{g}_{j-1}^{(n)}(\lambda_1, \cdots ,\lambda_n)
\end{align}
where $\vec{C}=(C_{21},C_{22})$ and $\vec{\xi}=(0,1,-1,0)$. 
$C_{21}$ and $C_{22}$ are 1-particle creation operators
and $D_{21}$ is 2-particle creation operator.  
$\hat{g}_j$ is a operator defined recursively,
\begin{align}
&\hat{g}_{j-1}^{(n)}(\lambda_1, \cdots ,\lambda_n)
=\dfrac{a^+(\lambda_{j-1},\lambda_j)}{a^-(\lambda_{j-1},\lambda_j)}
\hat{g}_{j-2}^{(n)}
(\lambda_1, \cdots ,\lambda_j,\lambda_{j-1}, \cdots ,\lambda_n)
.\hat{r}_{j-1,j}(\lambda_{j-1},\lambda_j),
\\
&\hat{g_1^{(n)}}(\lambda_1, \cdots ,\lambda_n)
= - \dfrac{f(\lambda_1,\lambda_2)}{c^+(\lambda_1,\lambda_2)}
\Pi_{k=3}^n \dfrac{a^+(\lambda_k,\lambda_2)}{b^-(\lambda_k,\lambda_2)}.
\end{align}
The vacuum state ($|0 \rangle $) is
a fermion Fock space defined as,
\begin{align}
|0\rangle =\prod_{j=1}^L |00\rangle_j, \ \ 
\rm{where} \ \
c_{j\sigma}|00\rangle_j=0 \ \ (\sigma=\uparrow, \downarrow),
\label{vacuumestate}
\end{align}
The main point of this algebraic construction
is its exchange property ,
\begin{align}
\vec{\Phi}_n(\lambda_1, \cdots ,\lambda_n)
= \dfrac{a^-(\lambda_{j-1},\lambda_j)}{a^+(\lambda_{j-1},\lambda_j)}
\vec{\Phi}_n(\lambda_1, \cdots ,\lambda_j,\lambda_{j-1}, \cdots ,\lambda_n)
.\hat{r}_{j-1,j}(\lambda_{j-1},\lambda_j),
\label{connpst}
\end{align}
which simplifies this construction.
After a long patient calculations (see Appendix B)
using the algebraic relations (\ref{relation1})--(\ref{relation2}),
we get the eigenvalue of transfer matrix
\begin{align}
&\Lambda(u,\{\lambda_l\}) = 
(a^+(u))^L \Pi_{j=1}^n (- \dfrac{a^-(u,\lambda_j)}{b^-(u,\lambda_j)})
+ (c^+(u))^L \Pi_{j=1}^n (-\dfrac{b^+(u,\lambda_j)}{c^+(u,\lambda_j)})
\nonumber\\
&
- (b^-(u))^L  \{
\Pi_{j=1}^n (-\dfrac{a^-(u,\lambda_j)}{b^-(u,\lambda_j)})
\Pi_{l=1}^m \frac{1}{\bar{b}(\mu_l,u)}
+\Pi_{j=1}^n (-\dfrac{b^+(u,\lambda_j)}{c^+(u,\lambda_j)})
\Pi_{l=1}^m \frac{1}{\bar{b}(u,\mu_l)}
\},
\label{eigenvalue.tm}
\end{align}
where 
$a^{\pm}(u)=a^{\pm}(u,0)$, $b^{\pm}(u)=b^{\pm}(u,0)$
and $c^{\pm}(u)=c^{\pm}(u,0)$.
And the Bethe Ansatz equations are,
\begin{align}
\Big( \dfrac{a^+(\lambda_j)}{b^-(\lambda_j)} \Big) ^L 
= \Pi_{l=1}^m \dfrac{1}{\bar{b}(\mu_l,\lambda_j)}, \ \
\Pi_{j=1}^n \bar{b}(\mu_l,\lambda_j)
= \Pi_{k \neq l}^m \dfrac{\bar{b}(\mu_l,\mu_k)}{\bar{b}(\mu_k,\mu_l)}.
\label{betheansatz}
\end{align}
To see the agreement with the coordinate one \cite{lieb},
we examine the energy spectrum and the momentum,
after the reparametrization 
$ e^{2h(\lambda_j)} /\tan \lambda_j = e^{{\rm i}k_j}$,
\begin{align}
E&= \frac{d}{d u}\log \Lambda(u,\{\lambda_l\}) |_{u=0}
= - 2 \Sigma_{j=1}^n \cos k_j +\dfrac{U}{4}(L-2n)
\label{energy}\\
P&= \rm{i} \log(\Lambda(u=0,\{\lambda_l\}))
= - \Sigma_{j=1}^n k_j.
\label{momentum}
\end{align}
Now we compare with the Ramos and Martins' results.
As we mentioned in \S 2, 
the initial condition of the transfer matrix
doesn't give the shift operator in the graded case.
Then the energy momentum is not given as (\ref{momentum}).
Beside the agreement of (\ref{betheansatz})(\ref{energy}),
(\ref{eigenvalue.tm}) differs from their results
by the overall factor. 
As is discussed in \cite{ramos},
the completeness can be discussed in the similar way \cite{essler1, essler2}.
Here the {\it regular} bethe ansats are the highest-weight states
of the $SO(4)$ symmetry. The other states are obtained
by the lowering operators.

Next we generalize the eigenvalue of the vacuum state
in the following way,
\begin{align}
&D_{11}(u)|0>=d_{11}(u)|0>, \ \
D_{22}(u)|0>=d_{22}(u)|0>, \\
&A_{11}(u)|0>=a_{11}a_{xx}(u)|0>, \ \
A_{22}(u)|0>=a_{22}a_{xx}(u)|0>, \ \
\end{align}
where $a_{11}$ and $a_{22}$ are constant.
The eigenvalue of transfer matrix $\Lambda(u,\{\lambda_l\})$
and the Bethe Ansatz equations are,
\begin{align}
&\Lambda(u,\{\lambda_l\}) = 
d_{22}(u) \Pi_{j=1}^n (- \dfrac{a^+(u,\lambda_j)}{b^-(u,\lambda_j)})
+ d_{11}(u) \Pi_{j=1}^n (-\dfrac{b^+(u,\lambda_j)}{c^+(u,\lambda_j)})
\nonumber\\
&
- a_{xx}(u)  \{
a_{11}\Pi_{j=1}^n (-\dfrac{a^-(u,\lambda_j)}{b^-(u,\lambda_j)})
\Pi_{l=1}^m \frac{1}{\bar{b}(\mu_l,u)}
+a_{22}\Pi_{j=1}^n (-\dfrac{b^+(u,\lambda_j)}{c^+(u,\lambda_j)})
\Pi_{l=1}^m \frac{1}{\bar{b}(u,\mu_l)}
\},
\label{generaleigenvalue}\\
&\Big( \dfrac{d_{22}(\lambda_j)}{a_{xx}(\lambda_j)} \Big) ^L 
= a_{11} \Pi_{l=1}^m \dfrac{1}{\bar{b}(\mu_l,\lambda_j)}, \ \
\Pi_{j=1}^n \bar{b}(\mu_l,\lambda_j)
=\frac{a_{11}}{a_{22}}
 \Pi_{k \neq l}^m \dfrac{\bar{b}(\mu_l,\mu_k)}{\bar{b}(\mu_k,\mu_l)}.
\label{generalbae}
\end{align}
Then we think about the twisted boundary condition,
\begin{align}
c_{L+1\uparrow}=e^{\rm{i} \theta_1} c_{1\uparrow}, \ \
c_{L+1\downarrow}=e^{\rm{i} \theta_2} c_{1\downarrow},
\label{twist}
\end{align}
where the transfer matrix is defined as
\begin{align}
&\tau^{\rm{twist}}(u)=\fstr_a \fT_a^{\rm{twist}}(u), 
\label{twisttm}
\\
&\fT_a^{\rm{twist}}(u) = \mathcal{V} \fT_a(u) \ \ \rm{where} \ \
\mathcal{V}
= e^{- \rm{i} \frac{\theta_1+\theta_2}{2} (n_{a\uparrow}+n_{a\downarrow}) 
+ \rm{i} \frac{\theta_2-\theta_1}{2} (n_{a\uparrow}-n_{a\downarrow}) }.
\end{align}
Because of the properties of the fermionic $R$-operator (\ref{eq.exp1}),
the global Yang Baxter relation (\ref{glybrel}) is satisfied 
with this monodromy matrix $\fT^{\rm{twist}}$
although the eigenvalues of its elements for the vacuum state
change to
\begin{align}
&d_{22}(u)= a^+(u)^L, \ \ 
d_{11}(u)= e^{\rm{i} \frac{\theta_1+\theta_2}{2} } c^+(u)^L, \\
&a_{11}= e^{- \rm{i} \theta_1 }, \ \
a_{22}= e^{- \rm{i} \theta_2 } , \ \
a_{xx}(u)= b^-(u)^L.
\end{align}
Here $\theta_1=\theta_2=0$
gives the PBC ($c_{j+L\sigma}=c_{j\sigma}$)
and $\theta_1=\theta_2=\pi$
gives the anti periodic boundary condition ($c_{j+L\sigma}=-c_{j\sigma}$).

\section{Algebraic Bethe Ansatz for the Quantum Transfer Matrix}
\setcounter{equation}{0}
\renewcommand{\theequation}{5.\arabic{equation}}
In this section,
we will diagonalize the quantum transfer matrix
in the same way as \S 4.
First we compare the quantum transfer matrix 
with the usual transfer matrix.
Then we make use of the similar algebraic structure
to diagonalize it.

In the row to row case where the transfer matrix, 
\begin{align}
\tau(u) = \fstr_a \fR_{aL}(u,0) \cdots \fR_{a1}(u,0)
=\tau(0)(1 + uH + O(u^2)),
\end{align}
is diagonalized,
we use the Yang Baxter equation to intertwine with respect to
auxiliary space, where $1 \rightarrow a$,
$2 \rightarrow b$ and  $3 \rightarrow j$ in (\ref{ybe}),
\begin{align}
\fR_{ab}(u_a,u_b) \fR_{aj}(u_a,u_j) \fR_{bj}(u_b,u_j)
= \fR_{bj}(u_b,u_j) \fR_{aj}(u_a,u_j) \fR_{ab}(u_a,u_b).
\end{align}

But in the case of the quantum transfer matrix,
it operates on the auxiliary space
and the monodromy matrix should be intertwined with respect to
the quantum space. We use the two equations,
\begin{align}
\fR_{jk}(u_j,u_k) \fR_{ak}(u_a,u_k) \fR_{aj}(u_a,u_j)
= \fR_{aj}(u_a,u_j) \fR_{ak}(u_a,u_k) \fR_{jk}(u_j,u_k),
\label{ybeqtm1}
\end{align}
\begin{align}
&\fR_{jk}(u_j,u_k)  \bar{\fR}_{ak}(u_k,u_a) \bar{\fR}_{aj}(u_j,u_a)
= \bar{\fR}_{aj}(u_j,u_a) \bar{\fR}_{ak}(u_k,u_a) \fR_{jk}(u_j,u_k),
\label{ybeqtm2}
\end{align}
which are equivalent to (\ref{ybe}).
In (\ref{ybeqtm1}), l.h.s. and r.h.s are reversed
and $1 \rightarrow a$, $2 \rightarrow j$ and $3 \rightarrow k$.
In (\ref{ybeqtm2}), we take the super transpose ($\fst$)
with respect to $3$ and  $1 \rightarrow j$, $2 \rightarrow k$
and $3 \rightarrow a$,
where the super transpose ($\fst$) is defined as 
$\fst_a=\fst_a^{\uparrow}\fst_a^{\downarrow}$ and
\begin{align}
&(X_1 n_{a \sigma} + X_2 c^{\dagger}_{a \sigma} 
+X_3 c_{a \sigma} + X_4 (1-n_{a \sigma}))^{\fst_a^{\sigma}}
\nonumber\\ \ \
&= X_1 n_{a \sigma} - X_2 c_{a \sigma} 
+X_3 c^{\dagger}_{a \sigma} + X_4 (1-n_{a \sigma}),
\ \ \ \ (\sigma=\uparrow, \downarrow)
\label{fstsigma}
\end{align} 
where $\sigma$ operators in space $a$ is expressed explicitly.
$\bar{\fR}_{aj}(u_j,u_a)$ can be expressed in the following way,
\begin{align}
& \bar{\fR}_{aj}(u_j,u_a)= \fR^{\fst_a}_{ja}(u_j,u_a)\\
& \ \ 
=\bar{\fR}^{(\uparrow)}_{aj}(u_j-u_a) 
\bar{\fR}^{(\downarrow)}_{aj}(u_j-u_a) 
+  \dfrac{\cos (u_j-u_a) }{\cos (u_j+u_a) } \tanh (h(u_j)-h(u_a)) \nonumber \\
& \ \ \ \
\times \bar{\fR}^{(\uparrow)}_{aj}(u_j+u_a)  
\bar{\fR}^{(\downarrow)}_{aj}(u_j+u_a) 
(2n_{j \uparrow}-1)(2n_{j \downarrow}-1),
\label{Rbar}\\
& 
\bar{\fR}^{(\sigma)}_{aj}(u)=\fR^{\sigma \fst_a^{\sigma}}_{ja}(u)
=a(u)(-n_{j \sigma}n_{a \sigma}+(1-n_{j \sigma})(1-n_{a \sigma}))
\nonumber\\
& \ \
-b(u)(n_{j \sigma}(1-n_{a \sigma})+(1-n_{j \sigma})n_{a \sigma})
+c(u)(c^{\dagger}_{j \sigma}c^{\dagger}_{a \sigma}+c_{j \sigma}c_{a \sigma}
).
\end{align}
Because of the properties for $\bar{\fR}_{aj}^{\sigma}(0)$,
\begin{align}
&\frac{\rm{d}}{\rm{d}u}\bar{\fR}_{aj+1}^{\sigma}(u)|_{u=0}
\bar{\fP}_{aj}^{\sigma} 
= H_{jj+1}^{\sigma}
\bar{\fP}_{aj+1}^{\sigma}\bar{\fP}_{aj}^{\sigma},
\\
&\bar{\fP}_{aj+1}^{\sigma}\bar{\fP}_{aj}^{\sigma} c_{j+1\sigma}
=c_{j\sigma} \bar{\fP}_{aj+1}^{\sigma}\bar{\fP}_{aj}^{\sigma},
\\
& \rm{where} \ \ \ \ 
\bar{\fP}_{aj}^{\sigma}=\bar{\fR}_{aj}^{\sigma}(0),
\ \
H_{jj+1}^{\sigma} 
= -(c_{j+1\sigma}^+c_{j\sigma}+c_{j\sigma}^+c_{j+1\sigma}),
\ \  (\sigma=\uparrow, \downarrow) 
\end{align}
we have the properties,
\begin{align}
&\bar{\tau}(u) = \fstr_a \bar{\fR}_{aL}(0,u) \cdots \bar{\fR}_{a1}(0,u)
=( 1 - uH + O(u^2))\bar{\tau}(0),\\
&\bar{\tau}(0) c_{j+1\sigma} = c_{j\sigma} \bar{\tau}(0).
\end{align}
Here $\bar{\tau}(0)$ is a right shift operator. Then
\begin{align}
&\tau(0)\bar{\tau(0)}=1,\\
&\tau(-u)\bar{\tau}(u) = 1 - 2u H + O(u^2).
\end{align}
Next we evaluate the partition function 
by using $\fR$ and $\bar{\fR}$,
\begin{align}
\ftr \rm{e}^{-\beta H} &= 
\lim_{N \rightarrow \infty}  \ftr(\tau(-u)\bar{\tau}(u))^{N/2}\\
&= \lim_{N \rightarrow \infty}  
\ftr \prod_{m=1}^{N/2}
\fstr_{a_{2m},a_{2m-1}}  
[ \fR_{a_{2m}L}(-u,0) \cdots \fR_{a_{2m}1}(-u,0)
\nonumber\\
&  \ \ \ \ \times
\bar{\fR}_{a_{2m-1}L}(u,0) \cdots \bar{\fR}_{a_{2m-1}1}(u,0) ]\\
&=\lim_{N \rightarrow \infty}  
\fstr \prod_{j}^L \ftr_j \fT^{QTM}_j(u,0),
\ \ \ \ (u=\frac{\beta}{N}) 
\label{partitionqtm}
\end{align}
where 
$N$ is called Trotter number which we
to be even
and
\begin{align}
&\fT^{QTM}_j(u,v) = \fR_{a_{N}j}(-u,v) \bar{\fR}_{a_{N-1}j}(v,u)
\cdots \fR_{a_{2}j}(-u,v) \bar{\fR}_{a_{1}j}(v,u),\\
&\tau^{QTM}(u,v)=\ftr_j {\fT^{QTM}_j}(u,v).
\label{qtm}
\end{align}
Here $\tau^{QTM}(u,v)$ is the QTM.
From (\ref{ybeqtm1})(\ref{ybeqtm2}), this monodromy operator
is intertwined as
\begin{align}
\fR_{jk}(u_j,u_k) \fT^{QTM}_{k}(u,u_k) \fT^{QTM}_{j}(u,u_j)
= \fT^{QTM}_{j}(u,u_j) \fT^{QTM}_{k}(u,u_k) \fR_{jk}(u_j,u_k).
\label{glybeqtm1}
\end{align}
After gauge transformation (\ref{gaugetrans.}), we have the property 
for the fermionic $R$-operator 
$\fR_{jk}(u_j,u_k)=\fR_{kj}(-u_k,-u_j)$.
Then we have
\begin{align}
\fR_{jk}(u_j,u_k) \fT^{QTM}_{j}(u,-u_j) \fT^{QTM}_{k}(u,-u_k)
= \fT^{QTM}_{k}(u,-u_k) \fT^{QTM}_{j}(u,-u_j) \fR_{jk}(u_j,u_k).
\label{glybeqtm2}
\end{align}
We see the similar relation 
for the usual transfer matrix (\ref{glybrel})(\ref{transfermatrix}).
And the QTM is diagonalized in the same way
except the generalization of the vacuum eigenvalue.
Next we introduce the magnetic field ($H = 2h$) 
and the chemical potential ($\mu$),
\begin{align}
H_{\rm{ext}} = -  \sum_{j=1}^L
\{ h (n_{j \uparrow}-n_{j \downarrow}) 
- \mu (n_{j \uparrow}+n_{j \downarrow}) \},
\end{align}
and evaluate the QTM algebraically.
$\ftr_j$ in (\ref{qtm}) is nothing but the anti periodic boundary condition 
($\theta_1=\theta_2=\pi$ in (\ref{twist}) and 
$\mathcal{V}^{\rm{apbc}}$ in (\ref{twisttm}) )
and the introduction of the magnetic field and the chemical potential 
is also applied to (\ref{twisttm}) as $\mathcal{V}^{\rm{m,c}}$,
\begin{align}
&\mathcal{V}= \mathcal{V}^{\rm{apbc}}  \mathcal{V}^{\rm{m,c}}, \\
&\mathcal{V}^{\rm{apbc}} 
= \rm{e}^{-\pi \rm{i} (n_{a\uparrow}+n_{a\downarrow})}, \ \
\mathcal{V}^{\rm{m,c}} 
= \rm{e}^{- \beta \mu (n_{a\uparrow}+n_{a\downarrow}) 
+ \beta h (n_{a\uparrow}-n_{a\downarrow}) }.
\end{align}
And the vacuum state ($|\Omega \rangle^{\rm{aux}}$) 
in the auxiliary Fock space
is defined as
\begin{align}
|\Omega \rangle^{\rm{aux}} 
= \prod_{m=1}^{N/2} |00\rangle_{a_{2m}} |11\rangle_{a_{2m-1}}, \ \
\rm{where} \ \
|11\rangle_a= c^{\dagger}_{a\uparrow} c^{\dagger}_{a\downarrow} |00\rangle_a
\end{align}
and $|00\rangle_a$ is given in (\ref{vacuumestate}).
Then $C_{21}$, $C_{22}$ and $D_{21}$ remain to be creation operators.
Then the generalized eigenvalue for the vacuum state 
is given by
\begin{align}
&d_{22}(v)=(a^+(-u,-v)c^+(-v,u))^{N/2}, \ \
d_{11}(v)= e^{-2 \beta \mu} (c^+(-u,-v)a^+(-v,u))^{N/2}, \\
&a_{11}= e^{\beta (-\mu + h)}, \ \ 
a_{22}= e^{\beta (-\mu - h)}, \ \
a_{xx}(v) = (-b^+(-u,-v)b^-(-v,u))^{N/2}.
\end{align}
And the eigenvalue and the bethe ansatz equations 
(\ref{generaleigenvalue})(\ref{generalbae}) are
\begin{align}
&\Lambda(u,\{\lambda_l\}) = 
(a^+(-u,-v)c^+(-v,u))^{N/2} 
\Pi_{j=1}^n (- \dfrac{a^+(u,\lambda_j)}{b^-(u,\lambda_j)})
\nonumber\\
&+e^{-2 \beta \mu} (c^+(-u,-v)a^+(-v,u))^{N/2} 
\Pi_{j=1}^n (-\dfrac{b^+(u,\lambda_j)}{c^+(u,\lambda_j)})
\nonumber\\
& - (-b^+(-u,-v)b^-(-v,u))^{N/2}  \{
e^{\beta (-\mu + h)}
\Pi_{j=1}^n (-\dfrac{a^-(u,\lambda_j)}{b^-(u,\lambda_j)})
\Pi_{l=1}^m \frac{1}{\bar{b}(\mu_l,u)}
\nonumber\\
&+ e^{\beta (-\mu - h)} 
\Pi_{j=1}^n (-\dfrac{b^+(u,\lambda_j)}{c^+(u,\lambda_j)})
\Pi_{l=1}^m \frac{1}{\bar{b}(u,\mu_l)}
\},\\
&\Big(- \dfrac{a^+(-u,-\lambda_j)c^+(-\lambda_j,u)}
{b^+(-u,-\lambda_j)b^-(-\lambda_j,u)} \Big) ^{N/2} 
= - e^{\beta(-\mu+h)} 
\Pi_{l=1}^m \dfrac{1}{\bar{b}(\mu_l,\lambda_j)}, \\
&\Pi_{j=1}^n \bar{b}(\mu_l,\lambda_j)
= e^{-2\beta h}
\Pi_{k \neq l}^m \dfrac{\bar{b}(\mu_l,\mu_k)}{\bar{b}(\mu_k,\mu_l)}.
\end{align}
When the system size is infinite ($L \rightarrow \infty$),
we see that the largest eigenvalue of the QTM with ($v=0$)
gives the partition function
because (\ref{partitionqtm}) can be expressed as 
a summation over the eigenstates ($\fstr$),
\begin{align}
\ftr \rm{e}^{-\beta H}
= \Lambda_1^L + (-1)^{p_2} \Lambda_2^L + (-1)^{p_3} \Lambda_3^L + \cdots,
\end{align}
where $p_l$ is a parity of $n$.
The largest eigenvalue ($\Lambda_1$)
is given by the sector ($n=N$ and $m=N/2$) and
this agrees with \cite{kluemper}
after the partial particle hole transformation.
Furthermore we can deal with the intrinsicly fermionic properties
which can not be expected in \cite{kluemper}.
The asymptotics of the correlation functions are given by 
another eigenvalue.
For example, the eigenvalue of ($n=N-1$ and $m=N/2-1$) sector 
gives the one particle Green function 
$<c^{\dagger}_{j\sigma} c_{j\sigma}>$. (see \cite{sakai})

\section{Concluding Remarks}

We studied the algebraic structure
for the 1D Hubbard model
in terms of the fermionic $R$-operator.
This approach is a powerful method when we treat fermion system directly.
The quantum transfer matrix (QTM) is naturally constructed.
Then it can be diagonalized 
just in the same way as that for the usual transfer matrix.
It has been a difficult problem 
and mapping to the spin system was a usual method,
where intrinsicly fermion properties can not be observed.
Our result can be applied to evaluate any correlation function
in addition to the bulk properties.

In this paper, we review the fermionic $R$-operator approach 
for the 1D Hubbard model
and show the supertrace and supertranspose for the fermion operator.
Then the algebraic aspect of this approach are discussed.
The properties of the $R$-operator itself
reflect the global properties of the model.
Then we construct the eigenstate algebraically
and diagonalize the transfer matrix.
There appears naturally the $r$-structure
which leads to the nested bethe ansatz equations.
The QTM is made from the fermionic $R$-operator 
and the supertrace of it.
By using the generalized algebraic bethe ansatz equations, 
we diagonalize the quantum transfer matrix (QTM).

For the future work, 
we will investigate more details about 
the solutions of this Bethe ansatz equations  for the QTM
to obtain the thermodynamic properties of this model.
Here the auxiliary space is finite and 
the Trotter limit ($N \rightarrow \infty$)
is desirable to obtain the exact result.
For the bulk properties, Juttner {\it et al.} \cite{kluemper}
obtained the nonlinear integral equations (NLIEs) to evaluate
the eigenvalue of the QTM,
where the infinite space can be easily considered.
We expect that the NLIEs for the another eigenvalue
can be the small deformation of the original one \cite{sakai}.
The one particle Green function $c^{\dagger}_{j\sigma} c_{k\sigma}$
and the spin correlation function $S^{+}_{j}S^{-}_k$
are our first problems, where we expect no singularity
with respect to the temperature.
The spin correlation function $S^{z}_{j}S^{z}_k$
is our next problem. We expect it to be more complicated.
When the Coulomb interaction is strong in the half-filed case,
it behave as the XXX Heisenberg chain, 
where the crossover phenomena in $S^{z}_{j}S^{z}_k$
is observed \cite{kluemper2}.

\vspace{5mm}
\begin{center}
{\bf Acknowledgment}
\end{center}
The authors are grateful to M. Shiroishi,
A. Kluemper and M. Wadati
for useful discussions and careful reading of the manuscript. 
This work 
is supported by Grant-in-Aid for 
JSPJ Fellows from the
Ministry of Education, Science, Sports and Culture of Japan. 

\vspace{30pt}
\begin{flushleft}
{\bf \Large
Appendix A; \hspace{1mm}
SO(4) symmetry}
\end{flushleft}
\setcounter{equation}{0}
\renewcommand{\theequation}{A.\arabic{equation}}

Here we will show some calculations which are omitted
in \S 3.
The global properties are obtained 
from the local properties of the fermionic $R$-operator.
As for the $SO(4)$ symmetry, it is
from (\ref{eq.exp1})--(\ref{eq.exp3}) to
(\ref{eq.trexp1})--(\ref{eq.trexp3}).
(\ref{eq.trexp1}) is proved from (\ref{eq.exp1}),
\begin{align}
\exp(- {\rm i} \theta S^{\alpha}) \tau(u,u_0) 
\exp( {\rm i} \theta S^{\alpha}) 
& =  {\rm Str}_a \prod_{j=1}^{L} {\rm e}^{- {\rm i} 
\theta S_j^{\alpha}} \fR_{aj}(u,u_0) 
{\rm e}^{{\rm i} \theta S_j^{\alpha}} \nonumber \\
& =  {\rm Str}_a \prod_{j=1}^{L} {\rm e}^{{\rm i} 
\theta S_a^{\alpha}} \fR_{aj}(u,u_0) 
{\rm e}^{- {\rm i} \theta S_a^{\alpha}}  \nonumber \\
& =  \tau(u,u_0).
\end{align}
In the same way, (\ref{eq.trexp2}) is obtained from (\ref{eq.exp2}). 
And (\ref{eq.trexp3}) is from (\ref{eq.exp3}), 
\begin{align}
&  \exp( {\rm i} \theta \eta^{\alpha}) \tau(u,u_0) 
\exp( {\rm i} \theta \eta^{\alpha}) \nonumber\\
&  = 
{\rm Str}_a \prod_{k=1}^{L/2} 
{\rm e}^{{\rm i} \theta \eta_{2k}^{\alpha}} \fR_{a,2k}(u,u_0) 
{\rm e}^{{\rm i} \theta \eta_{2k}^{\alpha}} 
{\rm e}^{- {\rm i} \theta \eta_{2k-1}^{\alpha}} \fR_{a,2k-1}(u,u_0) 
{\rm e}^{- {\rm i} \theta \eta_{2k-1}^{\alpha}} \nonumber \\
&  = {\rm Str}_a \prod_{k=1}^{L/2} 
{\rm e}^{{\rm i} \theta \eta_{a}^{\alpha}} \fR_{a,2k}(u,u_0) 
{\rm e}^{{\rm i} \theta \eta_{a}^{\alpha}}  
{\rm e}^{- {\rm i} \theta \eta_{a}^{\alpha}} \fR_{a,2k-1}(u,u_0) 
{\rm e}^{- {\rm i} \theta \eta_{a}^{\alpha}} \nonumber \\
&  = \tau(u,u_0). 
\end{align}
When we consider ${\tau^{-1}(u_0,u_0) \tau(u,u_0)}$,
it has $SO(4)$ symmetry (\ref{so4sym}).   

As for the invariance under the partial particle hole transformation,
we introduce ${\bar{V}}$ defined by
\begin{align}
  \bar{V} = \prod_{k=1}^{L/2}
\bar{V}_{2 k} V_{2k -1}
   = \bar{V}_{L} V_{L-1} \cdots \bar{V}_{2} V_1.
\end{align}
Now using (\ref{eq.ph1})(\ref{eq.ph2}), we obtain
\begin{align}
\bar{V}^{-1} \tau(u,u_0;-U) V 
& =  {\rm Str}_a \prod_{k=1}^{L/2}
\bar{V}_{2 k} {\mathcal{R}}^{\rm f}_{a,2k}(u,u_0;-U) V_{2 k} V_{2 k-1} 
{\mathcal{R}}^{\rm f}_{a,2k-1}(u,u_0) \bar{V}_{2k -1} \nonumber \\
& =  {\rm Str}_a \prod_{k=1}^{L/2} 
( - V_{a} {\mathcal{R}}^{\rm f}_{a,2k}(u,u_0;U) \bar{V}_{a}) 
( - \bar{V}_{a} {\mathcal{R}}^{\rm f}_{a,2k-1}(u,u_0) V_{a}) \nonumber \\
&= - \tau(u,u_0;U). \label{eq.trphtrans}
\end{align}
Here in the last equality we have used a property of the supertrace
\begin{equation}
{\rm Str}_{a} \left\{ V_a X V_a \right\} 
= - {\rm Str}_{a} \left\{ X  \right\}.
\end{equation}
for an arbitrary operator ${X}$.
The transformation law (\ref{eq.trphtrans}) means 
that the transfer matrix itself is not invariant 
under the partial particle-hole transformation (\ref{eq.pphtransf}). 
When we consider the product ${\tau^{-1}(u_0,u_0) \tau(u,u_0)}$,
as before, it is invariant (\ref{phinv}) under this transformation.

\begin{flushleft}
{\bf \Large
Appendix B; \hspace{1mm}
Fermionic $R$-operator 
and Algebraic relations among the elements of the Monodromy matrix}
\end{flushleft}
\setcounter{equation}{0}
\renewcommand{\theequation}{B.\arabic{equation}}

Here we show the simplified fermionic $R$-operator
and the algebraic relations among the elements of
the monodromy matrix $\fT$ which are necessary
when diagonalizing the transfer matrix.
The relations are obtained from the global Yang Baxter relation, 
\begin{align}
\fR_{ab}(u_a,u_b) \fT_{a}(u_a) \fT_{b}(u_b)
= \fT_{b}(u_b) \fT_{a}(u_a) \fR_{ab}(u_a,u_b)\\
\fT_a(u_a)=\fR_{aN}(u_a,0) \cdots \fR_{a1}(u_a,0).
\end{align}
Then some calculations that are omitted in \S 4
are supplemented.

We gauge transformed the fermionic $R$-operator (\ref{fermionicRhubbard}),
\begin{align}
I_a(u_a) I_b(u_b) \fR_{ab}(u_a,u_b) I_a(u_a)^{-1} I_b(u_b)^{-1}
\rightarrow
\fR_{ab}(u_a,u_b),
\label{gaugetrans.}
\end{align}
where
\begin{align}
I_a(u_a)= e^{h(u_a)(2n_{a\uparrow}-1)(2n_{b\downarrow}-1)},
\end{align}
to the simple form,
\begin{align} 
       \mathcal{R}_{12} (u_1,u_2) &= 
a^+(u_1,u_2) ( 
n_{1\uparrow} n_{1\downarrow} n_{2\uparrow} n_{2\downarrow}
 +   (1-n_{1\uparrow})(1-n_{1\downarrow})
                       (1-n_{2\uparrow})(1-n_{2\downarrow}) ) 
\nonumber\\&
- a^-(u_1,u_2) ( 
n_{1\uparrow} (1-n_{1\downarrow})
              n_{2\uparrow} (1-n_{2\downarrow}) 
+ (1-n_{1\uparrow}) n_{1\downarrow}
              (1-n_{2\uparrow}) n_{2\downarrow} ) 
\nonumber\\&
+ c^+(u_1,u_2) ( 
n_{1\uparrow} n_{1\downarrow}
              (1-n_{2\uparrow})(1-n_{2\downarrow}) 
+ (1-n_{1\uparrow}) (1-n_{1\downarrow})
               n_{2\uparrow} n_{2\downarrow} ) 
\nonumber\\&
+ c^-(u_1,u_2) ( 
(1-n_{1\uparrow}) n_{1\downarrow}
               n_{2\uparrow} (1-n_{2\downarrow}) 
+ n_{1\uparrow} (1-n_{1\downarrow})
               (1-n_{2\uparrow})n_{2\downarrow}) 
\nonumber\\&
+ b^+(u_1,u_2) ( 
(1-n_{1\uparrow}) (1-n_{1\downarrow}) 
                   - n_{1\uparrow} n_{1\downarrow} )
\nonumber\\& \hspace{5mm}
\times
( n_{2\uparrow}(1-n_{2\downarrow}) 
                   + (1-n_{2\uparrow})n_{2\downarrow} ) 
\nonumber\\&
+ b^-(u_1,u_2) ( 
n_{1\uparrow} (1-n_{1\downarrow}) 
                    + (1-n_{1\uparrow})n_{1\downarrow} )
\nonumber\\&\hspace{5mm}
\times
( (1-n_{2\uparrow})(1-n_{2\downarrow}) 
                    - n_{2\uparrow}n_{2\downarrow} ) 
\nonumber\\&
- d^+(u_1,u_2) (
 c^{\dagger}_{1\uparrow} c^{\dagger}_{1\downarrow}
                       c_{2\uparrow}c_{2\downarrow} 
+ c_{1\uparrow} c_{1\downarrow}
                 c^{\dagger}_{2\uparrow} c^{\dagger}_{2\downarrow} )  
\nonumber\\&
+ d^-(u_1,u_2) ( 
c^{\dagger}_{1\uparrow} c_{1\downarrow}
                 c_{2\uparrow} c^{\dagger}_{2\downarrow} 
+ c_{1\uparrow} c_{1\downarrow}
                 c^{\dagger}_{2\uparrow}c_{2\downarrow} )  
\nonumber\\&
+ e(u_1,u_2)  (   
( c^{\dagger}_{1\uparrow} c_{2\uparrow} 
                  + c^{\dagger}_{2\uparrow}c_{1\uparrow} )
( (1-n_{1\downarrow}) (1-n_{2\downarrow}) 
                  - n_{1\downarrow}n_{2\downarrow} ) 
\nonumber\\&\hspace{5mm}
+ ( c^{\dagger}_{1\downarrow} c_{2\downarrow}
                  + c^{\dagger}_{2\downarrow}c_{1\downarrow} )
( (1-n_{1\uparrow})(1-n_{2\uparrow}) 
                  - n_{1\uparrow} n_{2\uparrow} )    )       
\nonumber\\&
+ f(u_1,u_2)  (   
( c^{\dagger}_{1\uparrow} c_{2\uparrow} 
                  + c^{\dagger}_{2\uparrow} c_{1\uparrow} )
( n_{1\downarrow}(1-n_{2\downarrow}) 
                  + (1-n_{1\downarrow})n_{2\downarrow} ) 
\nonumber\\&\hspace{5mm}
+ ( c^{\dagger}_{1\downarrow} c_{2\downarrow} 
                  + c^{\dagger}_{2\downarrow} c_{1\downarrow} )
( n_{1\uparrow} (1-n_{2\uparrow}) 
                  + (1-n_{1\uparrow}) n_{2\uparrow} )   ),
\end{align}
where
\begin{align}
a^{\pm}(u_j,u_k) &= \cos^2 (u_j - u_k) 
\left[ 1 \pm \tanh \left\{ h(u_j) - h(u_k) \right\} 
\frac{\cos (u_j + u_k)}
{\cos (u_j - u_k)} \right], \nonumber \\
b^{\pm}(u_j,u_k) & =  
-\sin (u_j - u_k) \cos (u_j - u_k) 
\left[ 1 \pm \tanh 
\left\{ h(u_j) - h(u_k) \right\} 
\frac{\sin (u_j + u_k)}
{\sin (u_j - u_k)} \right] \nonumber \\
 & = - \sin (u_j - u_k) \cos (u_j - u_k) 
\left[ 1 \pm \tanh 
\left\{h(u_j) + h(u_k) \right\} 
 \frac{\cos (u_j + u_k)}
{\cos (u_j - u_k)} \right], \nonumber \\
c^{\pm}(u_j,u_k) & =  \sin^2 (u_j - u_k) 
\left[ 1 \pm \tanh \left\{ h(u_j) + h(u_k) \right\} 
\frac{\sin (u_j + u_k)}{\sin (u_j - u_k)} \right], \nonumber \\
d^{\pm}(u_j,u_k) & =  
1 \pm \tanh \left\{ h(u_j) - h(u_k) \right\} 
\frac{\cos (u_j - u_k)}{\cos (u_j + u_k)}, \nonumber \\
& =  1 \pm \tanh \left\{ h(u_j) + h(u_k) \right\} 
\frac{\sin (u_j - u_k)}{\sin (u_j + u_k)}, \nonumber  \\      
e(u_j,u_k)  & = 
\frac{\cos (u_j - u_k)}{\cosh \left\{ h(u_j) - h(u_k) \right\}},
\hspace{3mm}
f(u_j,u_k) =  \frac{\sin (u_j - u_k)}
{\cosh \left\{ h(u_j) + h(u_k) \right\}}.
\end{align}
We define the Monodromy operator as follows,
\begin{align}
\fT(u) &=
D_{11}n_{\uparrow}n_{\downarrow}
+D_{22}(1-n_{\uparrow})(1-n_{\downarrow})
+A_{11}n_{\uparrow}(1-n_{\downarrow})
+A_{22}(1-n_{\uparrow})n_{\downarrow}
\nonumber\\&
+D_{12}c^{\dagger}_{\uparrow}c^{\dagger}_{\downarrow}
+D_{21}c_{\uparrow}c_{\downarrow}
-A_{12}c^{\dagger}_{\uparrow}c_{\downarrow}
+A_{21}c_{\uparrow}c^{\dagger}_{\downarrow}
\nonumber\\&
+C_{11}n_{\uparrow}c^{\dagger}_{\downarrow}
+C_{12}c^{\dagger}_{\uparrow}n_{\downarrow}
-C_{21}c_{\uparrow}(1-n_{\downarrow})
+C_{22}(1-n_{\uparrow})c_{\downarrow}
\nonumber\\&
+B_{11}n_{\uparrow}c_{\downarrow}
-B_{12}c^{\dagger}_{\uparrow}(1-n_{\downarrow})
+B_{21}c_{\uparrow}n_{\downarrow}
+B_{22}(1-n_{\uparrow})c^{\dagger}_{\downarrow}.
\label{monodromy}
\end{align}
Note that the sign is a little tricky
as to get the simple r-structure
which is the most interesting point
of this story.
As the definition of the supertrace $\fstr$,
the transfer matrix is expressed as,
\begin{align}
\tau(u) &= \fstr_a \fT_a (u)
\label{fstr}\\
&= D_{11}(u)+D_{22}(u)-A_{11}(u)-A_{22}(u)
\label{tm}
\end{align}
Relations among $A_{11} \cdots D_{22}$ are
\begin{align}
C_{2a}(u) C_{2b}(u') 
&= \frac{a^-(u,u')}{a^+(u,u')} \hat{r}^{ab}_{cd}(u,u') C_{2c}(u') C_{2d}(u)
\label{relation1}
\nonumber\\
&+ \frac{f(u,u')}{c^+(u,u')} \xi_{ab} 
(D_{21}(u) D_{22}(u') - D_{21}(u') D_{22}(u)),
\\
A_{ab}(u) C_{2c}(u') 
&= - \frac{a^-(u,u')}{b^-(u,u')} 
\hat{r}^{bc}_{ed}(u,u') C_{2e}(u') A_{ad}(u)
+ \frac{e(u,u')}{b^-(u,u')} C_{2b}(u) A_{ac}(u')
\nonumber\\
&+ \xi_{bc} \frac{f(u,u')}{c^+(u,u')}
\{ B_{a1}(u) D_{22}(u') - \frac{e(u,u')}{b^-(u,u')}
 D_{21}(u) B_{a2}(u') 
\nonumber\\ 
&+ \frac{a^+(u,u')}{b^-(u,u')} D_{21}(u') B_{a2}(u) \},
\label{relA_abC}
\\
D_{11}(u) C_{2a}(u') 
&= - \frac{b^+(u,u')}{c^+(u,u')} C_{2a}(u') D_{11}(u)
- \frac{e(u,u')}{c^+(u,u')} D_{21}(u') C_{1a}(u)
\nonumber\\
&+ \frac{d^+(u,u')}{c^+(u,u')} D_{21}(u) C_{1a}(u')
\nonumber\\
&+ \frac{f(u,u')}{c^+(u,u')} \{ B_{11}(u) A_{2a}(u')
- B_{21}(u) A_{1a}(u') \} ,
\\
D_{22}(u') C_{2a}(u)
&= \frac{a^+(u,u')}{b^-(u,u')} C_{2a}(u) D_{22}(u')
- \frac{e(u,u')}{b^-(u,u')} C_{2a}(u') D_{22}(u),
\end{align}
\begin{align}
A_{ab}(u) D_{21}(u')
&= - D_{21}(u') A_{ab}(u)
\nonumber\\ 
&- \frac{e(u,u')}{b^-(u,u')} 
(C_{2b}(u') B_{a1}(u) - C_{2b}(u) B_{a1}(u')),
\\
D_{11}(u) D_{21}(u')
&= - \frac{d^+(u,u')}{c^+(u,u')} D_{21}(u) D_{11}(u')
+ \frac{a^+(u,u')}{c^+(u,u')} D_{21}(u') D_{11}(u)
\nonumber\\
&+ \frac{f(u,u')}{c^+(u,u')} \xi_{ab} B_{a1}(u) B_{b1}(u'),
\\
D_{22}(u') D_{21}(u)
&= - \frac{d^+(u,u')}{c^+(u,u')} D_{21}(u') D_{22}(u)
+ \frac{a^+(u,u')}{c^+(u,u')} D_{21}(u) D_{22}(u')
\nonumber\\
&- \frac{f(u,u')}{c^+(u,u')} \xi_{ab} C_{2a}(u') C_{2b}(u),
\end{align}
\begin{align}
B_{a1}(u) C_{2b}(u')
&= \frac{b^+(u,u')}{b^-(u,u')} C_{2b}(u') B_{a1}(u)
\nonumber\\
&+ \frac{e(u,u')}{b^-(u,u')}
(D_{21}(u') A_{ab}(u) - D_{21}(u) A_{ab}(u')),
\\
B_{a2}(u) C_{2b}(u') 
&= - \frac{b^+(u,u')}{b^-(u,u')} C_{2b}(u') B_{a2}(u)
\nonumber\\
&- \frac{e(u,u')}{b^-(u,u')} 
(D_{22}(u') A_{ab}(u) - D_{22}(u) A_{ab}(u')),
\\
C_{1a}(u) C_{2b}(u') 
&= - \frac{d^+(u,u')}{c^+(u,u')} C_{2a}(u) C_{1b}(u')
\nonumber\\
&+ \frac{f(u,u')}{c^+(u,u')} 
( A_{1a}(u) A_{2b}(u') - A_{2a}(u) A_{1b}(u'))
\nonumber\\
&+ \frac{a^-(u,u')}{c^+(u,u')} x^{ab}_{cd}(u,u') C_{2c}(u') C_{1d}(u)
\nonumber\\
&- \frac{f(u,u')}{c^+(u,u')} \xi_{ab} 
(D_{22}(u') D_{11}(u) - D_{21}(u') D_{12}(u)),
\label{relation2}
\end{align}
where
\begin{align}
&\hat{r} =
   \begin{pmatrix} 
        1 & & & \\
         & \bar{a} & \bar{b} & \\
         & \bar{b} & \bar{a} & \\
         & & & 1
   \end{pmatrix}, \ \  \ \ 
\hat{x} =
    \begin{pmatrix}
        1 & & & \\
         & \frac{d^-}{a^-} & -\frac{c^-}{a^-} & \\
         & -\frac{c^-}{a^-} & \frac{d^-}{a^-} & \\
         & & & 1
   \end{pmatrix},  
\\
&  
\bar{b}(u,v) = \dfrac{b^+(u,v) b^-(u,v)}{a^-(u,v) c^+(u,v)},
\hspace{3mm}
\bar{a}(u,v) = 1 - \bar{b}(u,v)
\end{align}

By using these relations, we operate
the transfer matrix (\ref{tm})
to the eigenstate defined by (\ref{eigenvecter}).
Here show the elements 
separately
\begin{align}
&D_{22}(u) |\Phi_n(\lambda_1, \cdots ,\lambda_n) \rangle
\nonumber\\&
= (a^+(u))^L  \Pi_{j=1}^n \dfrac{a^+(\lambda_j,u)}{b^-(\lambda_j,u)}
|\Phi_n(\lambda_1, \cdots ,\lambda_n) \rangle
\nonumber\\&
+\Sigma_{j=1}^n (a^+(\lambda_j))^L
|\Psi_{n-1}^{(1)}(u,\lambda_j; \{ \lambda_l \}) \rangle
\nonumber\\&
+\Sigma_{j=2}^n \Sigma_{l=1}^{j-1} H_1(u,\lambda_l,\lambda_j)
(a^+(\lambda_l)a^+(\lambda_j))^L
|\Psi_{n-1}^{(3)}(u,\lambda_j,\lambda_l; \{ \lambda_k \}) \rangle,
\label{d_22phi}
\end{align}
\begin{align}
&D_{11}(u)|\Phi_n(\lambda_1, \cdots ,\lambda_n) \rangle
\nonumber\\&
= (c^+(u))^L \Pi_{j=1}^n (- \dfrac{b^+(u,\lambda_j)}{c^+(u,\lambda_j)})
|\Phi_n(\lambda_1, \cdots ,\lambda_n) \rangle
\nonumber\\&
+\Sigma_{j=1}^n (b^-(\lambda_j))^L 
\Lambda^{(1)}(u=\lambda_j,\{ \lambda_l \})
|\Psi_{n-1}^{(2)}(u,\lambda_j; \{ \lambda_l \}) \rangle
\nonumber\\&
+\Sigma_{j=2}^n \Sigma_{l=1}^{j-1} H_2(u,\lambda_l,\lambda_j)
(b^-(\lambda_l) b^-(\lambda_j))^L
\nonumber\\& \times
\Lambda^{(1)}(u=\lambda_j,\{ \lambda_k \})
\Lambda^{(1)}(u=\lambda_l,\{ \lambda_k \})
|\Psi_{n-1}^{(3)}(u,\lambda_j,\lambda_l; \{ \lambda_k \}) \rangle,
\label{d_11phi}
\end{align}
\begin{align}
&\Sigma_a A_{aa}(u)
| \Phi_n(\lambda_1, \cdots ,\lambda_n) \rangle
\nonumber\\&
=(b^-(u))^L \Pi_{j=1}^n (-\dfrac{a^-(u,\lambda_j)}{b^-(u,\lambda_j)})
\Lambda^{(1)}(u,\{ \lambda_l \})
|\Phi_n(\lambda_1, \cdots ,\lambda_n) \rangle
\nonumber\\&
+\Sigma_{j=1}^n (b^-(\lambda_j))^L
\Lambda^{(1)}(u=\lambda_j,\{ \lambda_l \})
|\Psi_{n-1}^{(1)}(u,\lambda_j; \{ \lambda_l \}) \rangle
\nonumber\\&
+\Sigma_{j=1}^n (a^+(\lambda_j))^L
|\Psi_{n-1}^{(2)}(u,\lambda_j; \{ \lambda_l \}) \rangle
\nonumber\\&
+\Sigma_{j=2}^n \Sigma_{l=1}^{j-1} H_3(u,\lambda_l,\lambda_j)
(\bar b(\lambda_l,\lambda_j)-\bar a(\lambda_l,\lambda_j))
(a^+(\lambda_l)b^-(\lambda_j))^L
\nonumber\\& \times 
\Lambda^{(1)}(u=\lambda_j,\{ \lambda_k \})
|\Psi_{n-1}^{(3)}(u,\lambda_j,\lambda_l; \{ \lambda_k \}) \rangle
\nonumber\\&
+\Sigma_{j=2}^n \Sigma_{l=1}^{j-1} H_4(u,\lambda_l,\lambda_j)
(\bar b(\lambda_j,\lambda_l)-\bar a(\lambda_j,\lambda_l))
(a^+(\lambda_j)b^-(\lambda_l))^L
\nonumber\\& \times 
\Lambda^{(1)}(u=\lambda_l,\{ \lambda_k \})
|\Psi_{n-1}^{(3)}(u,\lambda_j,\lambda_l; \{ \lambda_k \}) \rangle,
\label{a_aaphi}
\end{align}
where
\begin{align}
&|\Psi_{n-1}^{(1)}(u,\lambda_j; \{ \lambda_l \}) \rangle
= - \dfrac{e(\lambda_j,u)}{b^-(\lambda_j,u)}
\Pi_{k \neq j}^n \dfrac{a^+(\lambda_k,\lambda_j)}{b^-(\lambda_k,\lambda_j)}
\vec{C}(u)
\nonumber\\
&\hspace*{3cm}
\otimes
\vec{\Phi_{n-1}}
(\lambda_1, \cdots ,\check{\lambda}_j, \cdots \lambda_n)
.\hat{O}_j^{(1)}(\lambda_j;\{ \lambda_k\}) . \vec{\mathcal{F}}
| 0 \rangle,
\nonumber\\
&|\Psi_{n-1}^{(2)}(u,\lambda_j; \{ \lambda_l \}) \rangle
=\dfrac{f(u,\lambda_j)}{c^+(u,\lambda_j)}
\Pi_{k \neq j}^n (-\dfrac{a^-(\lambda_k,\lambda_j)}{b^-(\lambda_k,\lambda_j)})
[\vec{\xi}.(\vec{B}(u) \otimes 1)] 
\nonumber\\
&\hspace*{3cm}
\otimes
\vec{\Phi}_{n-1}(\lambda_1, \cdots ,\check{\lambda}_j, \cdots ,\lambda_n)
.\hat{O}_j^{(1)}(\lambda_j;\{\lambda_k\}).\vec{\mathcal{F}}
| 0 \rangle,
\nonumber\\
&|\Psi_{n-1}^{(3)}(u,\lambda_j,\lambda_l; \{ \lambda_k \}) \rangle
=\Pi_{k \neq j,l}^n 
\dfrac{a^+(\lambda_k,\lambda_j)}{b^-(\lambda_k,\lambda_j)}
\dfrac{a^+(\lambda_k,\lambda_l)}{b^-(\lambda_k,\lambda_l)}
D_{21}(u) \vec{\xi} 
\nonumber\\
&\hspace*{3cm}
\otimes 
\vec{\Phi}_{n-2} (\lambda_1, \cdots, \check{\lambda}_j ,\cdots , 
\check{\lambda}_l ,\cdots, \lambda_n)
.\hat{O}_{lj}^{(2)}(\lambda_l,\lambda_j;\{ \lambda_k \})
.\vec{\mathcal{F}} | 0 \rangle,\\
& \ \
\hat{O}_j^{(1)}(\lambda_j;\{\lambda_k\})
= \Pi_{k=1}^{j-1} \dfrac{a^-(\lambda_k,\lambda_j)}{a^+(\lambda_k,\lambda_j)}
\hat{r}_{k,k+1}(\lambda_k,\lambda_j),\\
& \ \
\hat{O}_{lj}^{(2)}(\lambda_l, \lambda_j;\{\lambda_k\})
= \Pi_{k=1}^{l-1} \dfrac{a^-(\lambda_k,\lambda_j)}{a^+(\lambda_k,\lambda_j)}
\hat{r}_{k+1,k+2}(\lambda_k,\lambda_j)
\nonumber\\
& \ \ \ \
\times
\Pi_{k=l+1}^{j-1} \dfrac{a^-(\lambda_k,\lambda_j)}{a^+(\lambda_k,\lambda_j)}
\hat{r}_{k,k+1}(\lambda_k,\lambda_j)
\Pi_{k=1}^{l-1} \dfrac{a^-(\lambda_k,\lambda_j)}{a^+(\lambda_k,\lambda_j)}
\hat{r}_{k,k+1}(\lambda_k,\lambda_j),
\end{align}
\begin{align}
&H1(x,y,z)=\dfrac{a^+(y,x)e(z,x)f(y,x)}{b^-(y,x)b^-(z,x)c^+(y,x)}
+ \dfrac{d^+(y,x)f(y,z)}{c^+(y,x)c^+(y,z)},
\\
&H2(x,y,z)=\dfrac{d^+(x,y)f(y,z)}{c^+(x,y)c^+(y,z)}
- \dfrac{e(x,y)b^+(x,z)}{c^+(x,y)c^+(x,z)},
\\
&H3(x,y,z)=\dfrac{f(x,y)e(x,y)e(y,z)}{c^+(x,y)b^-(x,y)b^-(y,z)}
- \dfrac{a^+(x,y)e(x,z)f(x,y)}{b^-(x,y)b^-(x,z)c^+(x,y)},
\\
&H4(x,y,z)= -\dfrac{f(x,y)e(x,y)e(y,z)}{c^+(x,y)b^-(x,y)b^-(y,z)}
- 2 \dfrac{e(y,x)e(x,y)f(y,z)}{b^+(y,x)b^-(x,y)c^+(y,z)}
\nonumber\\& \ \
- \dfrac{a^-(x,y)f(x,z)e(x,y)}{b^-(x,y)c^+(x,z)b^+(x,y)}(1+\bar{a}(x,y)),
\end{align}
using the identities,
\begin{align}
&\vec{\Phi}_n(\lambda_1, \cdots ,\lambda_n)
=\vec{\Phi}_n(\lambda_j, \lambda_1,
 \cdots ,\check{\lambda}_j, \cdots  ,\lambda_n)
.\hat{O}_j^{(1)}(\lambda_j;\{\lambda_k\}),
\\
&\vec{\Phi}_n(\lambda_1, \cdots ,\lambda_n)
=\vec{\Phi}_n(\lambda_l, \lambda_j, \lambda_1, 
 \cdots ,\check{\lambda}_l, \cdots ,\check{\lambda}_j,\cdots ,\lambda_n)
.\hat{O}_{lj}^{(2)}(\lambda_l, \lambda_j;\{\lambda_k\}),\\
&H1(x,y,z)+H2(x,y,z)
\nonumber\\
& \ \
=( \bar{b}(y,z)-\bar{a}(y,z)) H3(x,y,z) 
+ (\bar{b}(z,y)-\bar{a}(z,y)) H4(x,y,z).
\end{align}

The first term of r.h.s in (\ref{a_aaphi})
is obtained by using $r$-structure of the (\ref{relA_abC})
and the following calculation.
$\mathcal{F}$ in (\ref{eigenvecter}) is fixed as follows,
\begin{align}
&\tr_a \mathcal{T}^{(1)}_a(u,\{\lambda_l\}).\vec{\mathcal{F}} 
= \Lambda^{(1)}(u,\{\lambda_l\}) \vec{\mathcal{F}},
\nonumber\\
&\hspace*{1cm}
\mathcal{T}^{(1)}_a(u,\{\lambda_l\})=\mathcal{L}^{(1)}_{an}(u,\lambda_n) 
\cdots \mathcal{L}^{(1)}_{a1}(u,\lambda_1),
\nonumber\\
&\hspace*{3cm}
\mathcal{L}^{(1)}(u,\lambda_j) = 
\begin{pmatrix}
1&0&0&0\\
0&\bar{b}(u,\lambda_j)&\bar{a}(u,\lambda_j)&0\\
0&\bar{a}(u,\lambda_j)&\bar{b}(u,\lambda_j)&0\\
0&0&0&1
\end{pmatrix}.
\end{align}
This L matrix $ \mathcal{L}^{(1)}$
is a permutation of the $r$-matrix.
After the reparametrization, 
\begin{align}
&\bar{a}(\lambda,\mu) = - \dfrac{U/2}
{\tilde{\mu} - \tilde{\lambda} - U/2 },
\hspace{3mm}
\bar{b}(\lambda,\mu) = 
\dfrac{\tilde{\mu} - \tilde{\lambda}}
{\tilde{\mu} - \tilde{\lambda} - U/2},
\nonumber\\
& \tilde{\mu} =\frac{ z^-(\mu) - 1/z^-(\mu)}{2} + \dfrac{U}{4},
\hspace{3mm}
z^-(\mu) =  - \exp(2h(\mu))/ \tan \mu,
\end{align}
we see that this is equivalent to 
the problem of diagonalizating the transfer matrix
for the XXX model and we know the solution,
\begin{align}
&\Lambda^{(1)}(u,\{\lambda_j\},\{\mu_l\})
= \Pi_{l=1}^m \frac{1}{\bar{b}(\mu_l,u)}
+\Pi_{j=1}^n \bar{b}(u,\lambda_j)
\Pi_{l=1}^m \frac{1}{\bar{b}(u,\mu_l)},\\
& \rm{where} \ \
\Pi_{j=1}^n \bar{b}(\mu_l,\lambda_j)
= \Pi_{k \neq l}^m \frac{\bar{b}(\mu_l,\mu_k)}{\bar{b}(\mu_k,\mu_l)}.
\end{align}
Now the eigenvalue of the transfer matrix is obtained,
\begin{align}
&\Lambda(u,\{\lambda_l\}) = 
(a^+(u))^L \Pi_{j=1}^n \dfrac{a^+(\lambda_j,u)}{b^-(\lambda_j,u)}
+ (c^+(u))^L \Pi_{j=1}^n (-\dfrac{b^+(u,\lambda_j)}{c^+(u,\lambda_j)})
\nonumber\\
&\hspace*{1cm}
- (b^-(u))^L \Pi_{j=1}^n (-\dfrac{a^-(u,\lambda_j)}{b^-(u,\lambda_j)})
\Lambda^{(1)}(u,\{\lambda_l\}).
\end{align}
And the unwanted terms in (\ref{d_22phi}), (\ref{d_11phi}) and (\ref{a_aaphi})
give the another constraint,
\begin{align}
&(\dfrac{a^+(\lambda_j)}{b^-(\lambda_j)})^L
= \Lambda^{(1)}(u=\lambda_j,\{\lambda_l\}).
\end{align}

\end{document}